\documentstyle[sprocl,epsf]{article}
\def\be{\begin{equation}}
\def\ee{\end{equation}}
\def\bea{\begin{eqnarray}}
\def\eea{\end{eqnarray}}
\newcounter{teile}
\def\eqnar{\stepcounter{equation}
\setcounter{teile}{\value{equation}} \setcounter{equation}{0}
\renewcommand\theequation{\arabic{teile}\alph{equation}}
\begin{eqnarray}}
\def\endeqnar{\end{eqnarray}
\renewcommand\theequation{\arabic{equation}}
\setcounter{equation}{\value{teile}}\hspace*{-2ex}}
\newcommand{\braket}[2]{\left\langle{#1}\right.\left|{#2}\right\rangle}
\newcommand{\bramket}[3]{\left\langle\,{#1}\,\left|\,{#2}\,
            \right|\,{#3}\,\right\rangle}
\newcommand{\intp}[1]{\int\frac{d^{3}{#1}}{(2\pi)^{3}}}
\newcommand{\intpp}[1]{\int\frac{d^{4}{#1}}{(2\pi)^{4}}}
\newcommand{\bqn}{\begin{eqnarray}}
\newcommand{\eqn}{\end{eqnarray}}
\newcommand{\nn}{\nonumber\\}
\newcommand{\ov}{\overline}
\newcommand{\eps}{\varepsilon}
\newcommand{\matrx}[4]{ \left( \begin{array}{*{2}{c}}
                                      #1 & #2 \\
                                      #3 & #4 
                                    \end{array} \right) }

\newcommand{\zweivek}[2]{ \left( \begin{array}{c} #1 \\ #2 
                               \end{array} \right) }

\begin{document}

\title{Nucleon and $\Delta$ in a covariant quark-diquark model}
\author{Volker Keiner}
\address{Institut f\"ur Theoretische Kernphysik,\\
         Universit\"at Bonn, Nussallee 14-16, D-53115 Bonn, FRG}
%\date{1996}
\maketitle
%%%%%%%%%%%%%%%%%%%%%%%%%%%%%%%%%%%%%%%%%%%%%%%%%%%%%%%%%%%%%%%%%%%%%%%%
\begin{abstract}
We develop a formally covariant quark-diquark model of 
the nucleon. The nucleon is treated as a bound state of a constituent
quark and a diquark interacting via a quark exchange. We include both
scalar and axial-vector diquarks. The underlying Bethe-Salpeter 
equation is transformed into a pair of coupled Salpeter equations. 
The space- and timelike electromagnetic form factors of the nucleon 
are calculated in the Mandelstam formalism for squared momentum transfers
up to $3 \; (\mbox{GeV/c})^2$ above threshold. The model is applied also
to the $\Delta(3/2)$. 
\end{abstract}
%%%%%%%%%%%%%%%%%%%%%%%%%%%%%%%%%%%%%%%%%%%%%%%%%%%%%%%%%%%%%%%%%%%%%%%%%
\section{Introduction} 
Many attempts have been made to describe the nucleon as a bound state
of a quark and a diquark. The most fundamental approach uses the
Nambu--Jona-Lasinio model \cite{ishii,huang,reinhardt} to
derive static properties of the nucleon and ground state baryons. 
Most diquark models favour the existence of two kinds of diquarks,
scalar and axial-vector ones. To
take into account the truly three quark structure and the
Pauli-principle a quark exchange between quark and diquark is assumed
to dominate the short range interaction \cite{reinhardt}. We adopt
these ideas in a constituent quark model based on the Bethe-Salpeter
equation. Following Salpeter \cite{salpeter} we assume an
instantaneous interaction and obtain a Salpeter-type equation. As the
only interaction we consider an (instantaneous) quark exchange  
between the quark and the diquark. Involving only scalar and 
axial-vector diquark channels we deduce a pair of coupled integral
equations similar to the framework of Ref. \cite{meyer}. Within a basis of
positive parity 
amplitudes with spin-1/2 we obtain a bound state solution 
for this equation by making use of
the Ritz variational principle. Then,
electromagnetic transition currents are calculated
using the Mandelstam formalism \cite{mandelstam}. For details of the
calculation see two recent papers \cite{kea,keb}.
We compute the electromagnetic form factors of the nucleon in both 
the space- and timelike regions. Extending the model to the 
$\Delta(3/2)$ resonance, the $N-\Delta$ transition form factors and 
the ratios $E2/M1$ and $C2/M1$ are calculated.   
%%%%%%%%%%%%%%%%%%%%%%%%%%%%%%%%%%%%%%%%%%%%%%%%%%%%%%%%%%%%%%%%%%%%%%%%
\section{The model}
\begin{figure}
\begin{center}
\leavevmode
\epsfxsize=0.75\textwidth
\epsffile{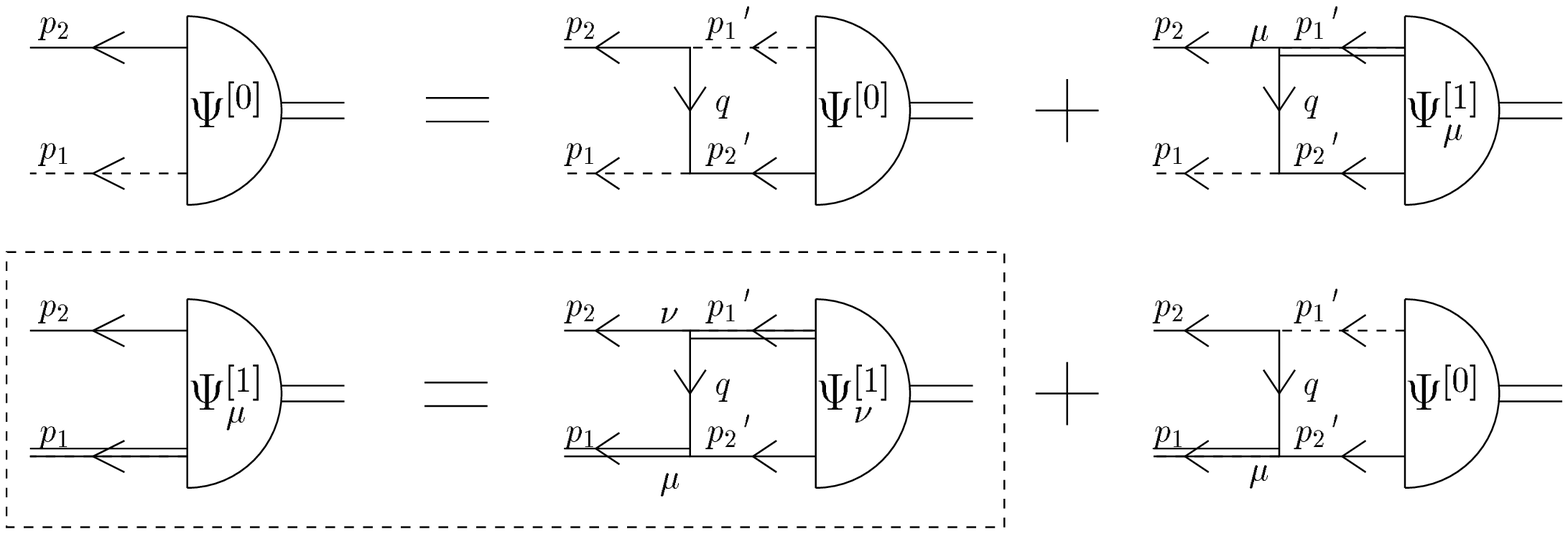}
\end{center}
\caption{The quark-diquark Salpeter equations. The framed part is the 
equation for the $\Delta(3/2)$ (see Sec. \protect\ref{secdelta}). }
\label{qdsalpeter}
\end{figure}

In momentum space the {\em Bethe-Salpeter amplitude} for a bound state
of a quark (index 2) and a scalar or axial-vector diquark (index 1) 
with total four momentum $P=p_1+p_2$ and relative momentum $p$ fulfills 
the {\em Bethe-Salpeter equation}: 
\bqn 
\label{BSequation}
\chi_P(p)_{(\mu)} = \Delta_1(p_1)_{(\mu)}^{\;\; (\nu)} \, S_2(p_2) \,
            \intpp{p'} \, \left( -i K(P,p,p') \, \chi_P(p') 
           \right)_{(\nu)} \; .
\eqn
$\Delta_1$ and $S_2$ are the propagators of the constituent diquark 
and the quark, respectively. For the v-diquark we choose 
$\Delta_{\mu \nu}=-ig_{\mu \nu}/(m_1^2-p_1^2+i\eps)$.
Following Salpeter \cite{salpeter} we neglect the time (i.e. energy)
dependence of the interaction kernel by 
assuming $K(P,p,p') = V(\vec p, \vec p\,')$ 
({\em instantaneous interaction}) in the rest frame of the bound
state.
Then, one can easily perform the $p^0$ integration. Defining in the rest
frame of the bound state the {\em Salpeter amplitude}
\bqn 
\label{salpeteramp}
\Phi(\vec p\,) := \left( \int \frac{d p^0}{2\pi} \, 
\chi_P(p^0,\vec p\,) \right)_{P=(M,\vec 0)}  \; ,
\eqn
one gets from Eq. (\ref{BSequation}) with standard techniques:
\bqn 
\label{salpeter1}
\Phi(\vec p\,) = \frac{1}{2 \omega_1} \left( 
 \frac{\Lambda_2^+(-\vec p\,) \gamma^0}{M-\omega_1-\omega_2}
+\frac{\Lambda_2^-(-\vec p\,) \gamma^0}{M+\omega_1+\omega_2}
\right)
\intp{p'} \, V(\vec p,\vec p\,') \, \Phi(\vec p\,')  \; , 
\eqn
with $\Lambda_2^\pm(-\vec p\,) = \frac{\omega_2 \pm H_2(-\vec p\,)}{2
  \omega_2}$  the usual Dirac projectors and
$\omega_i=\sqrt{\vec p_i\,^2+m_i^2}$. 
If we define 
\bqn
\Psi(\vec p\,) & := & \gamma^0 \, \Phi(\vec p\,) \;, \;\;\; 
W(\vec p, \vec p\,') := V(\vec p, \vec p\,') \, \gamma^0 \; ,
\eqn
we can rewrite Eq. (\ref{salpeter1}) in a Schr\"odinger-type
equation:
\bqn 
\label{salpeter2}
({\cal H} \, \Psi)(\vec p\,) & = & M \, \Psi(\vec p\,) \nn
& = & \frac{\omega_1+\omega_2}{\omega_2} H_2(\vec p\,) \, \Psi(\vec p\,)
      +\frac{1}{2 \omega_1} \, \intp{p'} \,
       W(\vec p, \vec p\,') \, \Psi(\vec p\,') \nn
& =: & ({\cal T} + {\cal V}) \, \Psi (\vec p\,) \; . 
\eqn
Of special importance in the Bethe-Salpeter approach is the correct
normalization of the amplitudes. This will guarantee the correct
behaviour of the form factors at $q^2 = 0$.
In the instantaneous approximation the usual {\em normalization
  condition} for the BS-amplitudes leads to
\bqn 
\label{normal}
\intp{p} \, (2 \omega_1) \, \ov \Phi(\vec p\,) \, \gamma^0 \, 
\Phi(\vec p\,) = 2 \, M \; ,
\eqn
with $\ov \Phi$ the adjoint Salpeter amplitude, fulfilling
$\ov \Phi(\vec p\,) = \Phi^\dagger(\vec p\,) \gamma^0$. 
This leads to the definition of a {\em scalar product} for 
$\Phi(\vec p\,)$: 
\bqn
\braket{\Phi_1}{\Phi_2} = \intp{p} \, (2 \omega_1) \, 
\ov\Phi_1(\vec p\,) \, \gamma^0 \, \Phi_2(\vec p\,) 
\; .  
\eqn
For $M \in {\cal R}^+$, the Hamiltonian ${\cal H}$
(Eq. (\ref{salpeter2})) 
has to be hermitian (and positive definite) with respect to the above
scalar product. In the instantaneous approximation the interaction
kernel is of the form 
\bqn
\label{interaction}
W(\vec p, \vec p\,') & \sim & 
-g^2 \, \frac{1}{\omega_q^2} \, (-\vec\gamma(\vec p+\vec p\,') + m_q)
\, \gamma^0 \; ,
\eqn
with $\omega_q$ the energy of the exchanged quark and
$g$ the dimensionless quark-diquark coupling parameter. To
obtain a stable solution of the Salpeter equation (\ref{salpeter2}) 
we have to introduce a form factor of the diquark. It is 
chosen to be of the form $\sim \mbox{exp}(-\lambda^2 k^2)$,
with $\lambda$ parametrizing the extension of the diquark. Transitions
between scalar and v-diquarks lead to 
a system of coupled Salpeter equations shown graphically in 
Fig. \ref{qdsalpeter}.
%%%%%%%%%%%%%%%%%%%%%%%%%%%%%%%%%%%%%%%%%%%%%%%%%%%%%%%%%%%%%%%%%%%%%
%%%%%%%%%%%%%%%%%%%%%%%%%%%%%%%%%%%%%%%%%%%%%%%%%%%%%%%%%%%%%%%%%%%%%%%%
%%%%%%%%%%%%%%%%%%%%%%%%%%%%%%%%%%%%%%%%%%%%%%%%%%%%%%%%%%%%%%%%%%%%%%%%
\section{Nucleon form factors}
\begin{figure}
\begin{center}
\leavevmode
\epsfxsize=0.7\textwidth
\epsffile{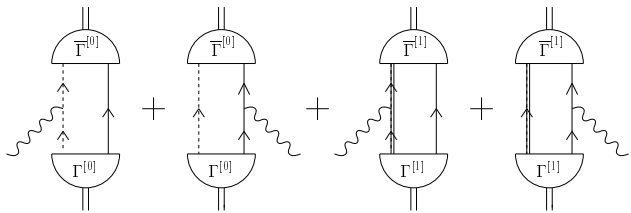}
\end{center}
\caption{The electromagnetic current is the sum of the diquark currents
  and the quark currents.}
\label{currentfig}
\end{figure}

\begin{figure}
\setlength{\unitlength}{1cm}
\begin{minipage}[t]{5.6cm}
\begin{picture}(5.6,3.5)
\epsfxsize=1.05\textwidth
\epsffile{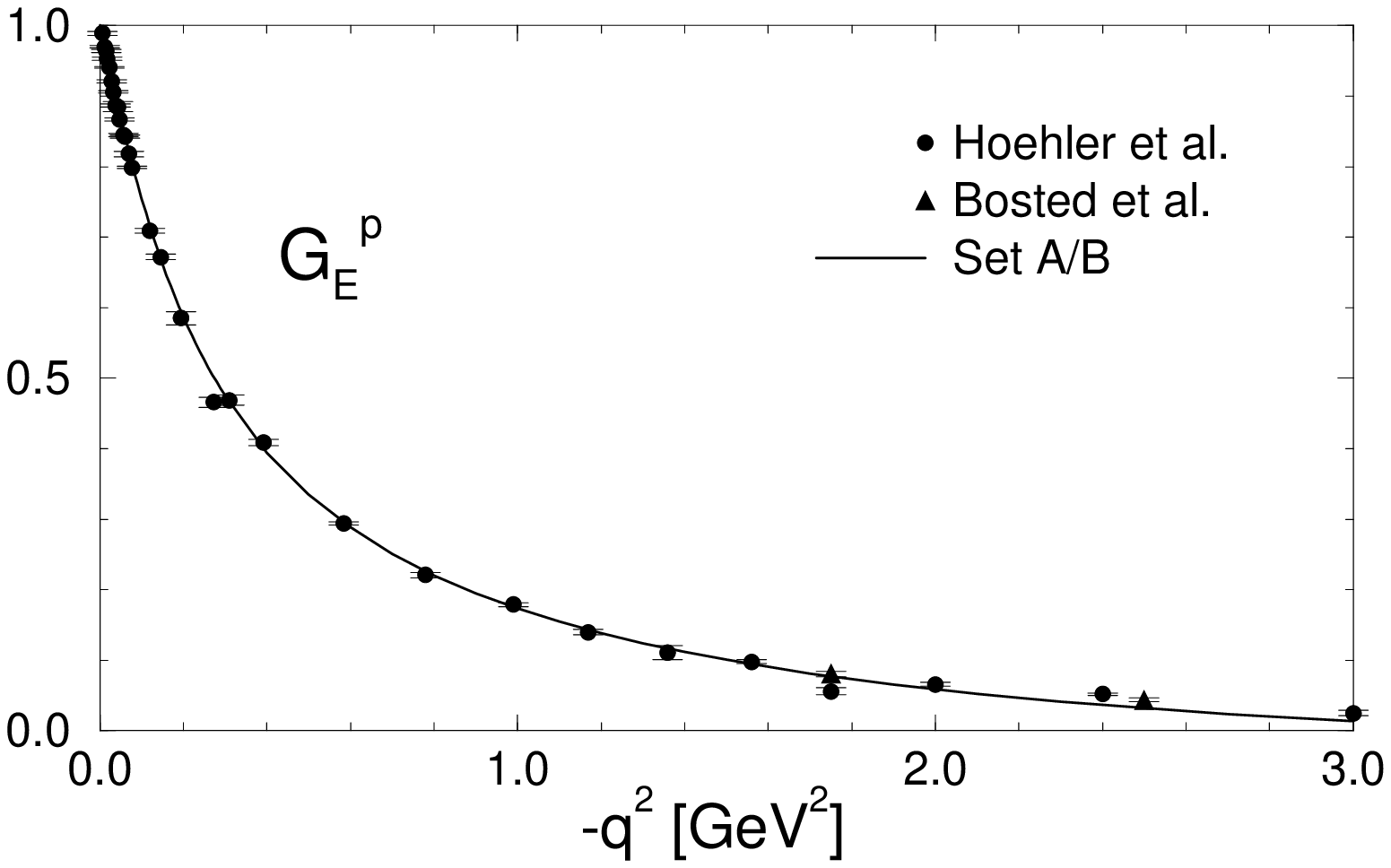}
\end{picture}
\end{minipage}
\hspace{0.1cm}
\begin{minipage}[t]{5.6cm}
\begin{picture}(5.6,3.5)
\epsfxsize=1.05\textwidth
\epsffile{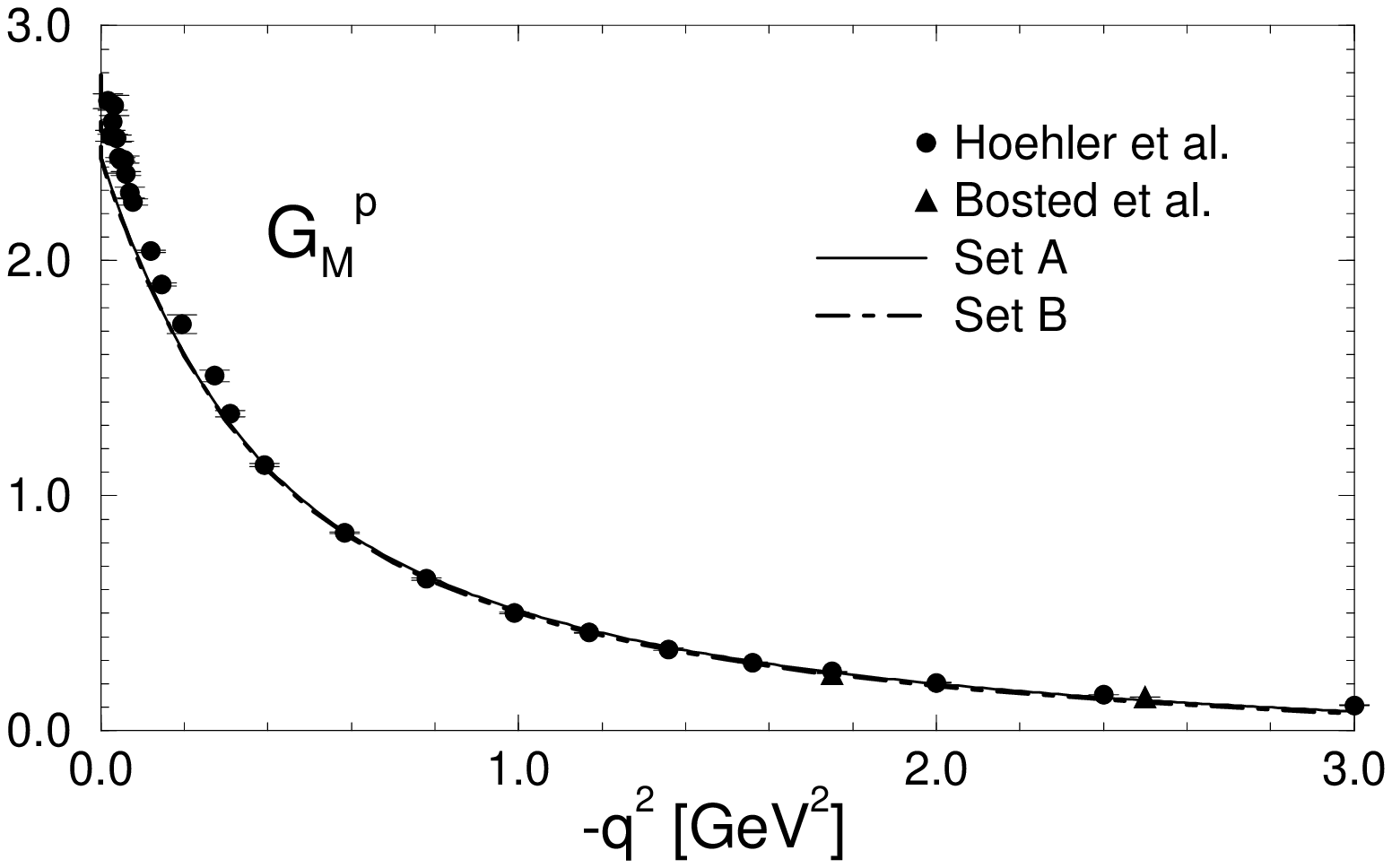}
\end{picture}
\end{minipage} \\
\begin{minipage}[t]{5.6cm}
\begin{picture}(5.6,3.5)
\epsfxsize=1.05\textwidth
\epsffile{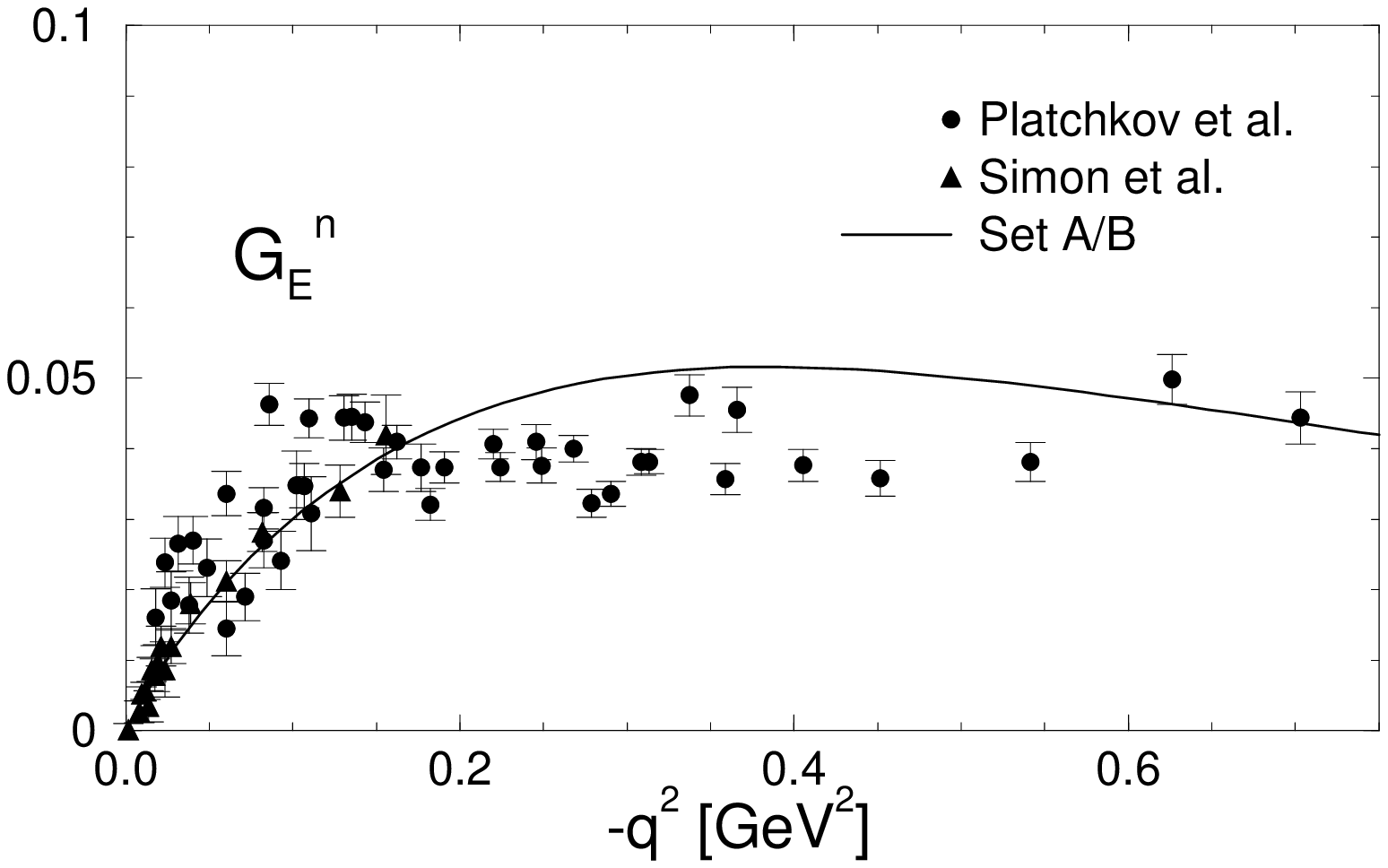}
\end{picture}
\end{minipage} 
\hspace{0.1cm}
\begin{minipage}[t]{5.6cm}
\begin{picture}(5.6,3.5)
\epsfxsize=1.05\textwidth
\epsffile{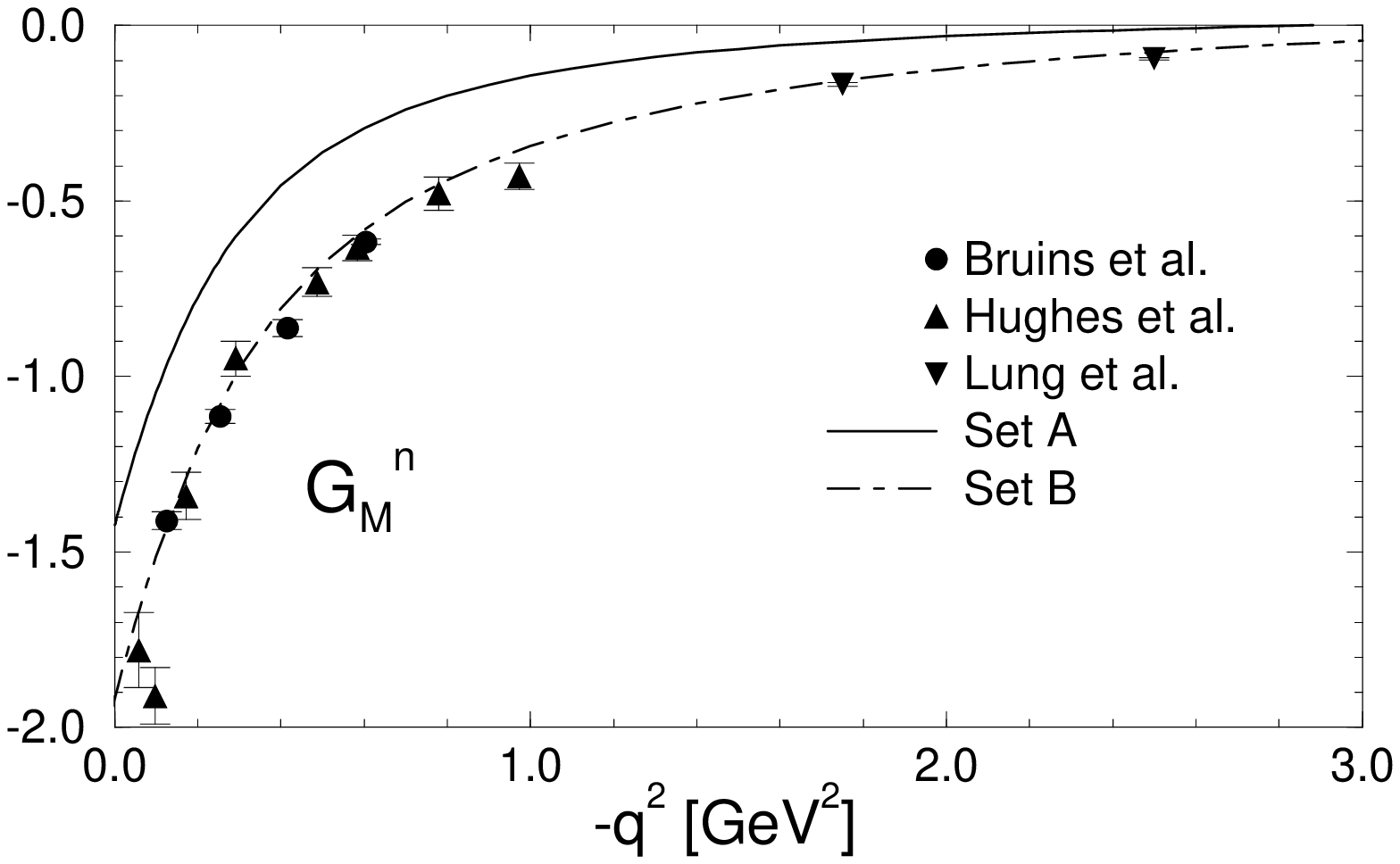}
\end{picture}
\end{minipage} 
\caption{The spacelike Sachs form factors of the nucleon. For the
  experimental data see the analysis of Ref. \protect\cite{ulf}. }
\label{spaceff}
\end{figure}

\begin{figure}
\begin{center}
\leavevmode
\epsfxsize=0.8\textwidth
\epsffile{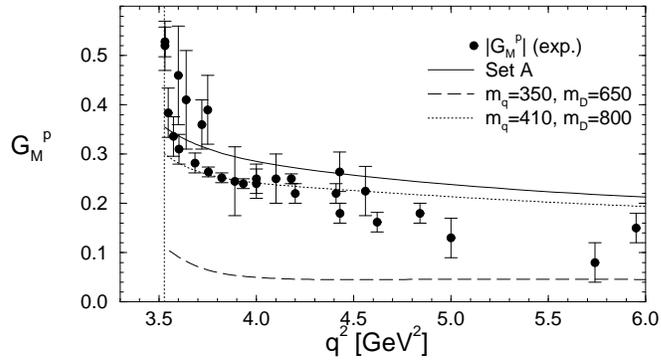}
\end{center}
\caption{The timelike magnetic form factor of the proton. 
  For the experimental data see Refs. \protect\cite{hammer,bini}. }
\label{timeff}
\end{figure}

\begin{figure}
\setlength{\unitlength}{1cm}
\begin{minipage}[t]{5.6cm}
\begin{picture}(5.6,3.5)
\epsfxsize=1.07\textwidth
\epsffile{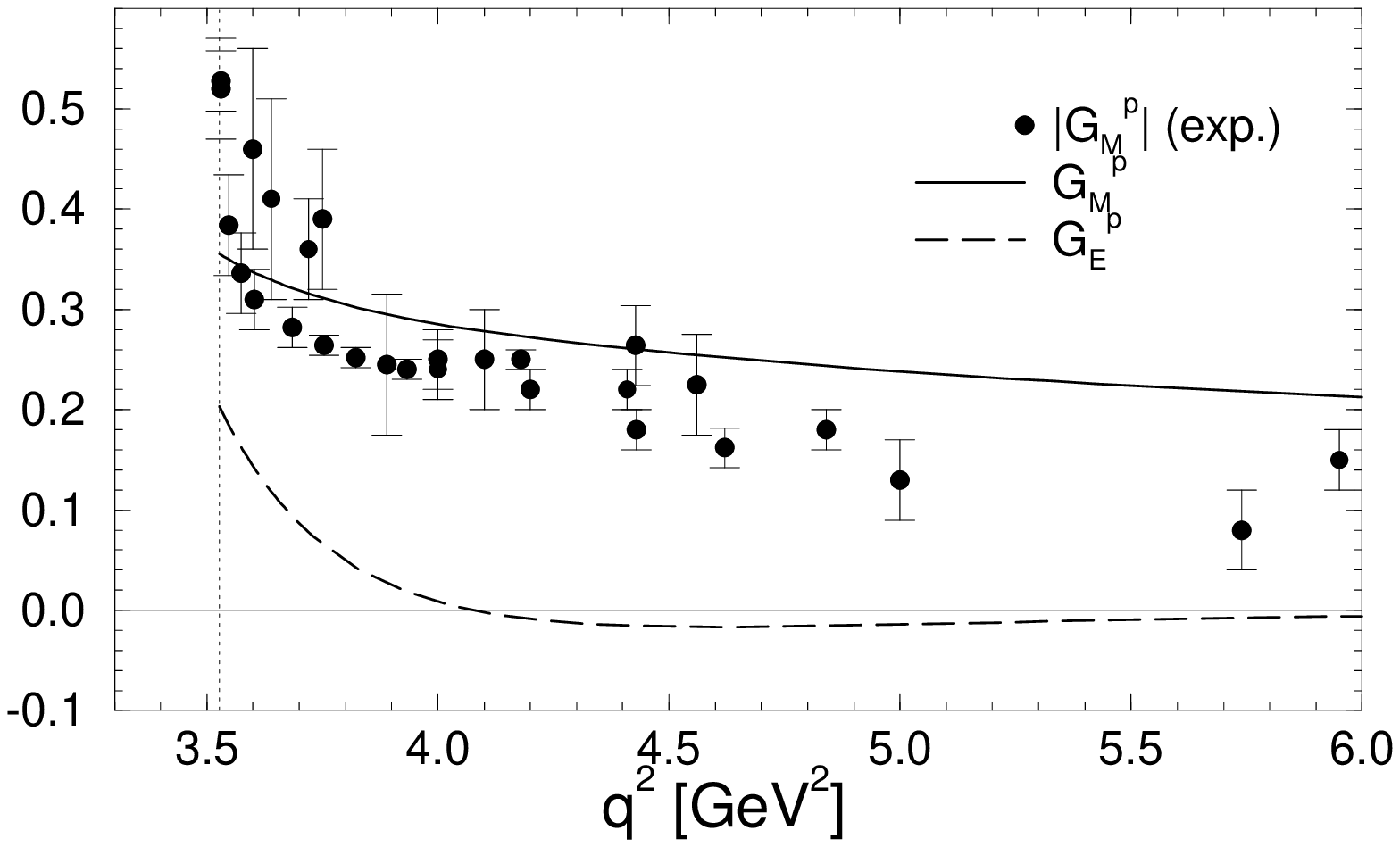}
\end{picture}
\end{minipage}
\hspace{0.2cm}
\begin{minipage}[t]{5.6cm}
\begin{picture}(5.6,3.5)
\epsfxsize=1.04\textwidth
\epsffile{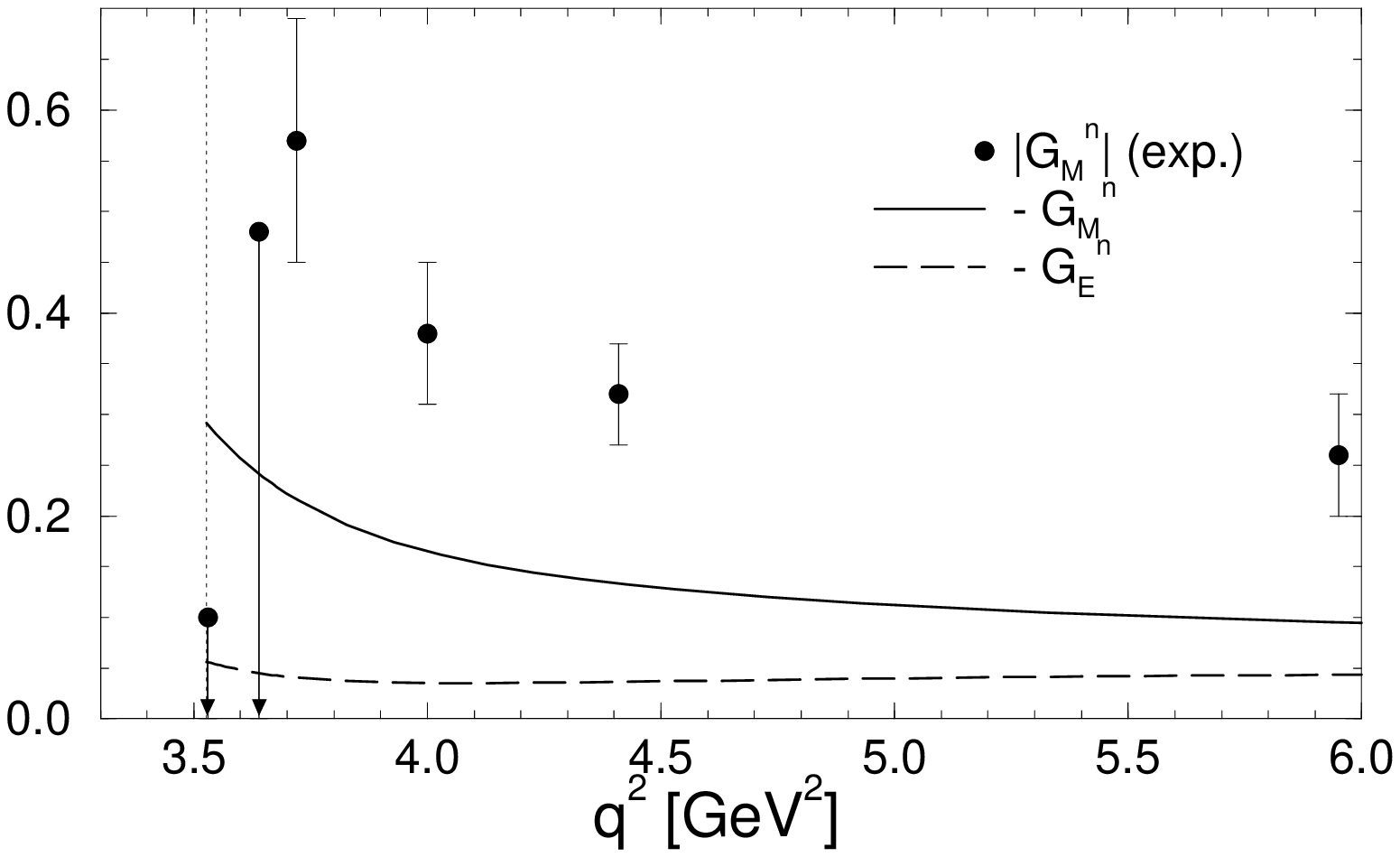}
\end{picture}
\end{minipage} 
\caption{The timelike magnetic and electric form factors of the proton
  (left panel) and of the neutron (right panel). For the experimental
  data see Refs. \protect\cite{hammer,bini}. }
\label{empnff}
\end{figure}

\begin{figure}
\begin{center} 
\leavevmode
\epsfxsize=0.70\textwidth
\epsffile{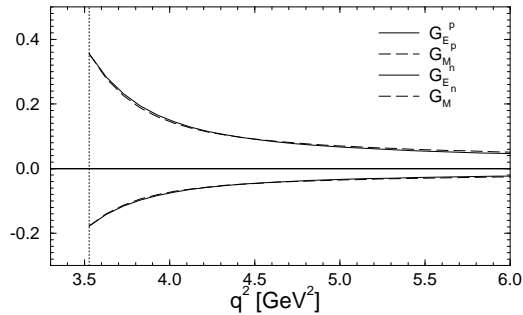}
\end{center}
\caption{The timelike nucleon form factors as they result from the quark
current with a scalar diquark as spectator {\em alone}. The upper two
curves are the proton form factors, the negative ones are those of the
neutron.}
\label{qunursff}
\end{figure}

The electromagnetic current is the sum of the diquark- and quark currents,
see Fig. \ref{currentfig}. 
%%%%%%%%%%%%%%%%%% Figure elm. current %%%%%%%%%%%%%%%%%%%%%%%%%%%%%%%%% 
%%%%%%%%%%%%%%%%%%%%%%%%%%%%%%%%%%%%%%%%%%%%%%%%%%%%%%%%%%%%%%%%%%%%%%%%
In the {\em Mandelstam formalism} one finds e.g. for the quark current:
\bqn 
\label{qcurrent}
\bramket{P's'}{j_\mu^{q}}{P s}&=&e_q\intpp{p} 
\ov\Gamma_{P'}(p')  
S_2^F(p_2') \gamma_\mu S_2^F(p_2)
\Gamma_P(p) \Delta_1^F(p_1)  ,
\eqn
with the vertex function  
$\Gamma_{P=(M,\vec 0\,)}(p)=\Gamma(\vec p \,)=
-i \intp{p'} V(\vec p, \vec p\,') \Phi(\vec p\,')$
in the instantaneous approximation. The form factors are obtained from:
\bqn
J_\mu:=e \, \ov u_{s'}(P') \left( \gamma_\mu \; F_1^N(q^2)
+\frac{i \sigma_{\mu \nu} q^\nu}{2 M} \, \kappa_N \, F_2^N(q^2) \right)  
u_s(P) \; ,
\eqn
with $q=P'-P$. The parameters used are given in Tab. \ref{param}. The scalar
%%%%%%%%%%%%%%%%%%%%%%%% Table Parameters %%%%%%%%%%%%%%%%%%%%%%%%%%%
\begin{table}[t]
\vspace{0.4cm}
\begin{center}
\begin{tabular}{|c||c|c|c|c|c|c|c|}
\hline
& $m_q$ & $m_S = m_V$ & $g^N$ & $g^\Delta$ & $\lambda$ & $\kappa_V$ & 
$\kappa_{SV}$ \\  
\hline
Set A &
440 MeV & 
800 MeV & 
17.76 &
8.50 &
0.30 fm &
1.1 & - \\
Set B &
440 MeV & 
800 MeV & 
17.76 &
8.50 &
0.30 fm &
-0.07 & 2.4  \\
\hline
\end{tabular}
\end{center}
\caption{The parameters of the model. }
\label{param}
\end{table}
%%%%%%%%%%%%%%%%%%%%%%%%%%%%%%%%%%%%%%%%%%%%%%%%%%%%%%%%%%%%%%%%%%%%%%%%%
and v-diquark parameters are chosen to be equal. $g^N$ is the quark-diquark 
coupling parameter for the nucleon (cf. Eq. (\ref{interaction})) and 
$g^\Delta$ that
for the $\Delta$ (see Sec. \ref{secdelta}). 
$\kappa_V$ is the anomalous magnetic
moment of the v-diquark introduced via
\bqn
j_{\mu; b a}^{V-V}\sim-(p + p')_\mu g_{b a} 
+(1+\kappa_V)(p_b-p'_b)g_{\mu a}+(1+\kappa_V)(p'_a-p_a)g_{\mu b} \, .
\eqn 
We compare the results of the parameter Set A with those of Set B 
where we allow also for photon-induced scalar--v-diquark transitions 
according to
\bqn
j_{\mu}^{S-V}&\sim&\frac{(1+\kappa_{SV})}{M_N} \, \eps_{\mu \nu \rho
  \lambda} \,   
\epsilon^\nu \, {p'}_2^\rho \, p_2^\lambda  \; .
\eqn 
Fig. \ref{spaceff} shows the Sachs form factors of the nucleon in the spacelike
%%%%%%%%%%%%%%%%%%%%%%%%%%%%%%%%%%%%%%%%%%%%%%%%%%%%%%%%%%%%%%%%%%%%%%%
%%%%%%%%%%%%%%%%%%%%%%%%%%%%%%%%%%%%%%%%%%%%%%%%%%%%%%%%%%%%%%%%%%%%%%
region. Note that $\kappa_V$ and $\kappa_{SV}$ only affect the spin-flip
currents, i.e. the magnetic form factors. We find a very good description 
of all form factors up to $-q^2=3 \; \mbox{GeV}^2$. The resulting static 
%%%%%%%%%%%%%%%%%%%%%%% Table Static nucleon properties %%%%%%%%%%%%%%%%%
\begin{table}[t]
\vspace{0.4cm}
\begin{center}
\begin{tabular}{|c||c|c|c|c|c|c|}
\hline
 & $\sqrt{\langle r^2 \rangle_E^p}$ & $\langle r^2 \rangle_E^n$ &
$\sqrt{\langle r^2 \rangle_M^p}$ &
$\sqrt{\langle r^2 \rangle_M^n}$ &
$\mu_p/\mu_N$ & $\mu_n/\mu_N$ \\ 
\hline
Set A & 0.79 fm  & -0.110 fm$^2$ & 
0.72 fm  & 0.86 fm  & 
2.45 & -1.42  \\ 
Set B & 0.79 fm  & -0.110 fm$^2$ &
0.74 fm  & 0.75 fm &
2.44 & -1.91  \\ 
exp. & 0.847 fm & -0.113 fm$^2$ & 
0.836 fm & 0.889 fm & 
2.793 & -1.913  \\
\hline
\end{tabular}
\end{center}
\caption{Static nucleon properties as they
  result from the threshold behaviour of the electromagnetic 
  nucleon form factors. For the experimental data see the analysis of 
  Ref. \protect\cite{ulf}. }
\label{stat}
\end{table}
%%%%%%%%%%%%%%%%%%%%%%%%%%%%%%%%%%%%%%%%%%%%%%%%%%%%%%%%%%%%%%%%%%%%%%%
properties are listed in Tab. \ref{stat}. The inclusion of scalar--v-diquark
transitions is a possibility to describe also the neutron magnetic form 
factor. Fig. \ref{timeff} shows the proton magnetic form factor in the 
%%%%%%%%%%%%%%%%%%%%%%%%%%%%%%%%%%%%%%%%%%%%%%%%%%%%%%%%%%%%%%%%%%%%%%%
%%%%%%%%%%%%%%%%%%%%%%%%%%%%%%%%%%%%%%%%%%%%%%%%%%%%%%%%%%%%%%%%%%%%%%%
timelike region. We find that the threshold value is very mass 
dependent, see the dashed and dotted curves. 
The dotted curve corresponds to a best fit 
to the electromagnetic form factors in the space- and timelike region. 
The masses, however, lie below the $\Delta(1232)$ threshold. The definition
of the Sachs form factors leads to the threshold conditions: 
\bqn
\label{thresh}
G_E(4 M^2)=G_M(4 M^2) \; \leftrightarrow \; 
J_+(4 M^2)=2 M \, \frac{d}{dP'} J_0(4 M^2) \; ,
\eqn
with $P'=|\vec P'|=q$, $P=(M,\vec 0\,)$ and $J_+=1/2 (J_1+i J_2)$. 
Fig. \ref{empnff} shows $G_{E,M}^{n,p}$. We find that
%%%%%%%%%%%%%%%%%%%%%%%%%%%%%%%%%%%%%%%%%%%%%%%%%%%%%%%%%%%%%%%%%%%%%%%
%%%%%%%%%%%%%%%%%%%%%%%%%%%%%%%%%%%%%%%%%%%%%%%%%%%%%%%%%%%%%%%%%%%%%%%
Eq. (\ref{thresh}) is violated in our model. 
This is due to the choice of the propagator
of the v-diquark. Considering the second diagram in 
Fig. \ref{currentfig} alone yields the result of Fig. \ref{qunursff}, which
%%%%%%%%%%%%%%%%%%%%%%%%%%%%%%%%%%%%%%%%%%%%%%%%%%%%%%%%%%%%%%%%%%%%%%%%%%
%%%%%%%%%%%%%%%%%%%%%%%%%%%%%%%%%%%%%%%%%%%%%%%%%%%%%%%%%%%%%%%%%%%%%%%%%%
shows that the relativistic treatment is correct. 
%%%%%%%%%%%%%%%%%%%%%%%%%%%%%%%%%%%%%%%%%%%%%%%%%%%%%%%%%%%%%%%%%%%%%%%%%%
\section{$N-\Delta$ transition form factors}
\label{secdelta}
\begin{figure}
\begin{center}
\leavevmode
\epsfxsize=0.7\textwidth
\epsffile{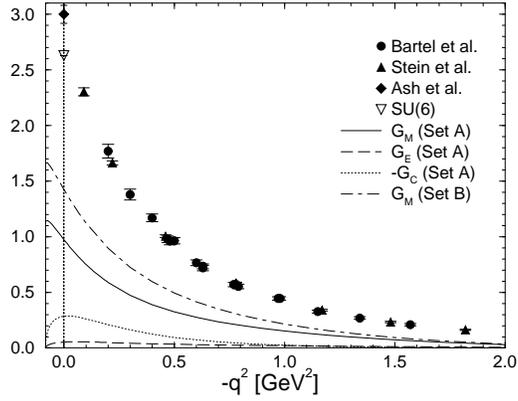}
\end{center}
\caption{The N-$\Delta$ transition form factors. }
\label{ndff}
\end{figure}

\begin{figure}
\begin{center} 
\leavevmode
\epsfxsize=0.57\textwidth
\epsffile{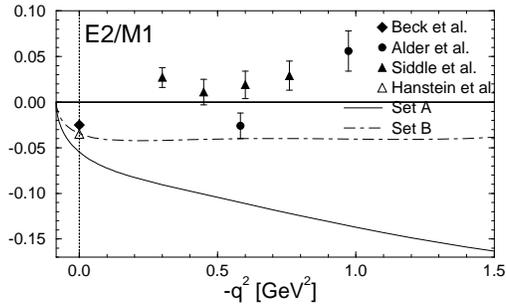}
\end{center}
\caption{The ratio $E2/M1$. }
\label{emratio}
\end{figure}

\begin{figure}
\begin{center} 
\leavevmode
\epsfxsize=0.55\textwidth
\epsffile{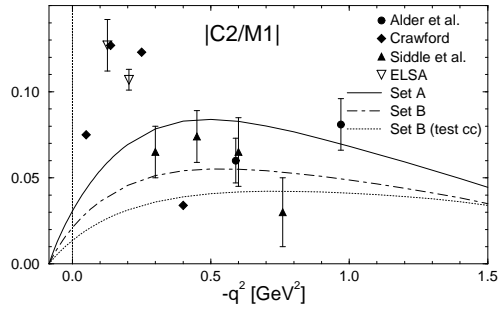}
\end{center}
\caption{The ratio $|C2/M1|$. }
\label{cmratio}
\end{figure}

In our model, the $\Delta(3/2)$ appears as a bound state of a v-diquark
and a quark, see the framed part in Fig. \ref{qdsalpeter}. 
To fix the $\Delta$ mass at 1232 MeV we introduce the quark-diquark 
coupling parameter $g^\Delta$, see Tab. \ref{param}. 
The $\Delta$ amplitude is normalized
according to $\braket{\Psi_\Delta}{\Psi_\Delta}=2 M_\Delta$. 
The electromagnetic $N-\Delta$ transition is decomposed via \cite{dev}
\bqn
e \, J_\mu(q^2)&=&e \, \sqrt{2/3} \; \ov u_{s'}^\beta(P') \, J_{\beta
  \mu} \, u_s(P) \; , \\
\mbox{with } \;\;\; J_{\beta \mu} &=& G_M(q^2) \, G_{\beta \mu}^M + 
G_E(q^2) \, G_{\beta \mu}^E + G_C(q^2) \, G_{\beta \mu}^C \; ,
\eqn
with $G_M, G_E, G_C$ the magnetic dipole, electric and Coulomb
quadrupole form factors, respectively. In the rest frame of the 
incoming nucleon this leads to: 
\bqn
\label{gme}
\zweivek{G_M(q^2)}{G_E(q^2)}&=& 
\sqrt{3/2} \;
\frac{2 \, \sqrt{2}}{g(q^2)} \,
\matrx{\frac{\sqrt{3}}{4}}{\frac{1}{12}}
     {\frac{1}{4 \sqrt{3}}}{-\frac{1}{12}} 
\zweivek{J_+(q^2)}{J'_+(q^2)} \\
\label{gc}
G_C(q^2) & = & - \sqrt{3/2} \;
\frac{4 M_\Delta M_N}{g(q^2) \, \sqrt{6 \, Q^+ \, Q^-}} \, J_0(q^2) \; ,  
\eqn
with $Q^\pm=(M_\Delta \pm M_N)^2-q^2$, 
$g(q^2)=((M_\Delta+M_N)/(2 M_N))\sqrt{Q^-}$ and 
$J'_+=\bramket{+3/2}{J_+}{+1/2}$. 
Fig. \ref{ndff} shows the calculated form factors and the experimental
%%%%%%%%%%%%%%%%%%%%%%%%%%%%%%%%%%%%%%%%%%%%%%%%%%%%%%%%%%%%%%%%%%%%%%%
%%%%%%%%%%%%%%%%%%%%%%%%%%%%%%%%%%%%%%%%%%%%%%%%%%%%%%%%%%%%%%%%%%%%%%%
$G_M$ \cite{kec}. We find that 
$G_M$ comes out a factor of 2 too low. Here, the inclusion of scalar to
v-diquark transitions leads only to a small improvement. Note that
$G_C$ is negative, and $G_E(q_s^2) \approx G_C(q_s^2) \approx 0$ at 
the pseudothreshold
$q_s^2=(M_\Delta-M_N)^2$. Figs. \ref{emratio} and \ref{cmratio} 
show the ratio 
%%%%%%%%%%%%%%%%%%%%%%%%%%%%%%%%%%%%%%%%%%%%%%%%%%%%%%%%%%%%%%%%%%%%%%%
%%%%%%%%%%%%%%%%%%%%%%%%%%%%%%%%%%%%%%%%%%%%%%%%%%%%%%%%%%%%%%%%%%%%%%%
$E2/M1$ and the absolute value $|C2/M1|$, respectively. 
With Set B we find the correct threshold value 
$E2/M1=-3.4 \; \%$ \cite{drechsel}. 
For $C2/M1$ we obtain a positive sign, with the absolute
value describing the data well. Note that we follow consistently
the definitions of Ref. \cite{dev}. The two lines of Set B indicate the
variance of the calculated ratio due to the fact that the $N-\Delta$
current is only partially conserved. 
%%%%%%%%%%%%%%%%%%%%%%%%%%%%%%%%%%%%%%%%%%%%%%%%%%%%%%%%%%%%%%%%%%%%%%
\section{Summary}
We developed a formally covariant quark-diquark model of the nucleon.
The nucleon is assumed to be composed of a quark and a
scalar/axial-vector diquark which interact via an instantaneous quark
exchange. 
With a single set of parameters we are able to describe the nucleon
electromagnetic form factors in the space- and timelike region for
momentum transfers up $3 \; \mbox{GeV}^2$ above threshold. In this
picture, the $\Delta(3/2)$ is a bound state of a quark and a
v-diquark. The results for the $N-\Delta$ transition agree only 
qualitatively with experiment, but we find the correct value for
$E2/M1$.

%%%%%%%%%%%%%%%%%%%%%%%%%%%%%%%%%%%%%%%%%%%%%%%%%%%%%%%%%%%%%%%%%%%%%%%
\section*{Acknowledgments}
I am indebted to H.R. Petry and B.C. Metsch. This work was
supported by the Deutsche Forschungsgemeinschaft.
%%%%%%%%%%%%%%%%%%%%%%%%%%%%%%%%%%%%%%%%%%%%%%%%%%%%%%%%%%%%%%%%%%%%%%%%

%%%%%%%%%%%%%%%%%%%%%%%%%%%%%%%%%%%%%%%%%%%%%%%%%%%%%%%%%%%%%%%%%%%%%%%%
\end{document}